\documentclass[psfig,showpacs,fleqn,nobibnotes]{revtex4}

\usepackage{amsmath}
\usepackage{graphicx}
\usepackage{float}
\usepackage{subfigure}

\def\lsim{\raise0.3ex\hbox{$<$\kern-0.75em\raise-1.1ex\hbox{$\sim$}}}
\def\gsim{\raise0.3ex\hbox{$>$\kern-0.75em\raise-1.1ex\hbox{$\sim$}}}

\def\beq{\begin{equation}}
\def\eeq{\end{equation}}
\def\bea{\begin{eqnarray}}
\def\eea{\end{eqnarray}}
\def\bq{\begin{quote}}
\def\eq{\end{quote}}

\newcommand{\rr}{\mbox{\boldmath $r$}}

\newcommand{\rp}{\mbox{\boldmath $p$}}

\def\gappeq{\mathrel{\rlap {\raise.5ex\hbox{$>$}}
{\lower.5ex\hbox{$\sim$}}}}

\def\lappeq{\mathrel{\rlap{\raise.5ex\hbox{$<$}}
{\lower.5ex\hbox{$\sim$}}}}

\def\Toprel#1\over#2{\mathrel{\mathop{#2}\limits^{#1}}}

\newcommand{\rk}{\mbox{\boldmath $k$}}

\begin{document}

\title{\bf Heavy  Quark Photoproduction in Proton - Proton Collisions}
\pacs{25.75.Dw, 13.60.Le}
\author{ V.P. Gon\c{c}alves $^{1}$, M.V.T. Machado  $^{2,\,3}$}
\affiliation{$^{1}$ \rm Instituto de F\'{\i}sica e Matem\'atica,  Universidade
Federal de Pelotas\\
Caixa Postal 354, CEP 96010-090, Pelotas, RS, Brazil\\
$^{2}$  \rm Universidade Estadual do Rio Grande do Sul - UERGS\\
 Unidade de Bento Gon\c{c}alves. CEP 95700-000. Bento Gon\c{c}alves, RS, Brazil\\
$^{3}$ \rm High Energy Physics Phenomenology Group, GFPAE,  IF-UFRGS \\
Caixa Postal 15051, CEP 91501-970, Porto Alegre, RS, Brazil}

\begin{abstract}
 We 
calculate the photoproduction of heavy quarks in
proton-proton collisions at RHIC, Tevatron and LHC energies,  where the photon reaches energies larger than those ones  accessible at DESY-HERA.  The integrated cross section
and the rapidity distributions for open charm and bottom production are computed employing sound high
energy QCD formalisms. For the linear pQCD approaches we consider both the  usual collinear factorization   and the $k_{\perp}$-factorization formalisms, whereas for the nonlinear QCD (saturation) calculations one considers the Golec-Biernat-W\"{u}sthoff  and the Iancu-Itakura-Munier parameterizations for the dipole cross section within the color dipole picture. 
\end{abstract}

\maketitle

\section{Introduction}

Heavy quark production in hard collisions of hadrons, leptons, and photons has been considered as a clean test of perturbative QCD (For a recent review see, e.g., Ref \cite{reviewhq}). This process provides not only many tests of perturbative QCD, but also some of the most important backgrounds to new physics processes, which have  motivated an extense phenomenology at DESY-HERA,  Tevatron and  LHC. These studies are mainly motivated by the strong dependence of the cross section on the behavior of the gluon distribution, which determines the QCD dynamics at high energies.
In particular, 
 the heavy quark photoproduction on
nucleon and nuclei targets has been studied in detail in Refs.
\cite{Mariottomagno,Goncalves:2003kp,Goncalves:2003zy}, considering the several available  scenarios for the QCD dynamics at high energies.
The results of those analysis show that  future  electron-proton (nucleus)
colliders at HERA and RHIC \cite{HERAeA,raju}, probably could
determine whether parton distributions saturate and a stringent constraint to 
the behavior of the gluon distribution in the full
kinematical range could be posed. Along these lines, in Ref. \cite{periferal3} we have analyzed the possibility of using
ultraperipheral heavy ion collisions (UPC's) as a photonuclear collider
and studied the heavy quark production  assuming distinct formalisms
for the QCD evolution.

Recently, Klein and Nystrand \cite{Kleinpp} have analyzed the quarkonium photoproduction in proton-proton collisions, considering the 
energetic protons as search of  large electromagnetic fields. 
In particular, these authors have used the photon spectrum from Ref. \cite{Dress} and a photon-proton cross section for quarkonium production obtained by fitting the H1 and ZEUS data. The main conclusion of that study is that the cross sections are large enough for this channel to be observed experimentally and  that this reaction can be used to study the gluon distribution in protons at small values of the Bjorken $x$ variable. This achievement  motivates the analysis of other processes that are sensitive to the gluon distribution and have larger cross sections. 

In this paper, we study the photoproduction of  heavy quarks in proton-proton collisions considering  distinct theoretical scenarios, which have been analyzed in detail in our previous papers \cite{Goncalves:2003kp,Goncalves:2003zy,periferal3}.  Our main motivation comes from the fact that in this process the photon reaches energies higher than those currently accessible at DESY - HERA. Similar
motivation is present in  Ref. \cite{Kleinpp}, where vector meson photoproduction is investigated.   Here, we estimate, for the first time, the total cross
section and the rapidity dependence of the photoproduction
of heavy quarks in proton-proton collisions,
considering the $k_{\perp}$-factorization approach taking into account  distinct unintegrated gluon distributions. Moreover, based on the color dipole picture \cite{dipole}, we compute the total cross section for charm and bottom photoproduction within the Iancu-Itakura-Munier (IIM) model for the dipole cross section \cite{iancu_munier}. These results will be used  as input for our further calculations in proton-proton collisions.   For comparison, we also present the predictions from the collinear factorization approach.

In relativistic heavy ion colliders, the heavy nuclei give rise to strong electromagnetic fields, which can interact with each other. In a similar way, these processes also occur when considering energetic protons in $pp(\bar{p})$ colliders. Namely, quasi-real photons scatters off protons at very high energies in the current hadron colliders. In particular, the heavy quark photoproduction cross section in a proton-proton collision is given by,
\begin{equation}
   \sigma(p+p \rightarrow p+X+Q\overline{Q}) = 2 \int_{0}^{\infty} \frac{dN_{\gamma}(\omega)}{d\omega}
   \, \sigma_{\gamma p\rightarrow Q\overline{Q} X}\left(W_{\gamma p}^2= 2\,\omega\,\sqrt{S_{NN}}  \right) \, d \omega \; , 
\label{eq:sigma_pp}
\end{equation}
where $\omega$ is the photon energy in the center-of-mass frame (c.m.s.), $W_{\gamma p}$ is the c.m.s. photon-proton energy and $\sqrt{S_{NN}}$ denotes the proton-proton c.m.s.energy.   The  photon spectrum is given by  \cite{Dress},
\begin{eqnarray}
\frac{dN_{\gamma}(\omega)}{d\omega} =  \frac{\alpha_{\mathrm{em}}}{2 \pi\, \omega} \left[ 1 + \left(1 - 
\frac{2\,\omega}{\sqrt{S_{NN}}}\right)^2 \right] 
\left( \ln{\Omega} - \frac{11}{6} + \frac{3}{\Omega}  - \frac{3}{2 \,\Omega^2} + \frac{1}{3 \,\Omega^3} \right) \,,
\label{eq:photon_spectrum}
\end{eqnarray}
with the notation $\Omega = 1 + [\,(0.71 \,\mathrm{GeV}^2)/Q_{\mathrm{min}}^2\,]$ and $Q_{\mathrm{min}}^2= \omega^2/[\,\gamma_L^2 \,(1-2\,\omega /\sqrt{S_{NN}})\,] \approx (\omega/
\gamma_L)^2$, where  $\gamma_L$ is the Lorentz factor. The expression above is derived considering the Weizs\"{a}cker-Williams method of virtual photons and using an elastic proton form factor (For more detail see Refs. \cite{Kleinpp,Dress}). It is important to emphasize that the expression (\ref{eq:photon_spectrum})  is based on a heuristic approximation,
which leads to an overestimation of the cross section at high energies ( $\approx 11 \%$ at $\sqrt{s}=1.3$ TeV)  in comparison with the more rigorous derivation of the photon spectrum for elastic scattering on protons derived in Ref. \cite{kniehl}.   For a more detailed comparison among the different photon spectra see Ref. \cite{nys_fluxo}. 

 Other process of interest for heavy quark production is the coherent interaction between two photons. As verified in Ref. \cite{per2} for ultraperipheral heavy ion collisions, the QCD dynamics implies an enhancement of the cross section in comparison with Quark - Parton Model calculations \cite{klein_vogt}. However, its  cross section is a factor about  $800$ smaller than for the one - photon process. Furthermore, differently from two-photon interactions, which leave both photon intact, photoproduction should only dissociate the proton target. Then one way to select photoproduction events is to eliminate events where both protons breakup and requiring a single rapidity gap.   

In what follows, we briefly present sound models for the heavy quark photoproduction at the photon level, $\sigma_{tot}\,(\gamma p \rightarrow Q\overline{Q} X)$,  which enter as input in the calculations for proton-proton collisions in Eq. (\ref{eq:sigma_pp}). Our results are compared with the DESY-HERA data \cite{h1data} and the prediction from the IIM model for this process is presented for the first time. In the last section, the numerical results for the rapidity $y$ of the produced states and their total cross sections are shown. A comparison on the order of magnitude of the cross sections for the distinct approaches is performed. Moreover, we present our main conclusions.

\section{A comparison among high energy approaches}

The photon-proton cross section can be calculated considering different theoretical scenarios \cite{smallx}.
Usually, one calculates it assuming the
validity of the collinear factorization, where the cross sections
involving incoming hadrons are given,  at all orders, by the
convolution of intrinsically non-perturbative, but universal,
quantities - the parton densities, with perturbatively calculable
hard matrix elements, which are process dependent. In this
approach,  all partons involved are assumed to be on mass shell,
carrying only longitudinal momenta, and their transverse momenta
are neglected in the QCD matrix elements. The heavy quark cross section is given in terms of the
convolution between the elementary cross section for the
subprocess $\gamma g \rightarrow Q \overline{Q}$ and the
probability of finding a gluon inside the proton, namely the
 gluon distribution. The photoproduction cross section at leading order is given by \cite{Gluck78},
\begin{eqnarray}
\sigma_{tot}\,(\gamma p\rightarrow Q\overline{Q}X) & = & 4\,\pi\,e_Q^2
\int_{4m_Q^2}^{W_{\gamma p}^2}\,\frac{dM_{Q\overline{Q}}^2}{M_{Q\overline{Q}}^2} \,\alpha_{em}\alpha_s(\mu_F^2)\,g_p(x,\mu_F^2)\,\nonumber \\
& \times &\, \left[
\left(1+\beta +\frac{1}{2}\beta^2\right)\ln\,\left(\frac{1+\sqrt{1-\beta}}
{1-\sqrt{1-\beta}}\right)  
- (1+\beta) \sqrt{1-\beta}\right] \,\,,
\label{sigpho}
\end{eqnarray}
where $M_{Q\overline{Q}}$ is the invariant
mass of the heavy quark  pair, with
$x=M^2_{Q\overline{Q}}/W_{\gamma p}^2$  and  $g_p(x, \mu_F^2)$
is the gluon density inside the proton at the factorization scale $\mu_F^2$. In addition,  $m_Q$ is  the heavy quark mass, $e_Q$ is its electric charge and
$\beta=4\,m_Q^2/M_{Q\overline{Q}}^2$. For the  present
purpose we will use $\mu_F^2=4\,m_Q^2$, with $m_c = 1.5$ GeV and
$m_b = 4.5$ GeV. In our further  calculations
on the collinear approach one takes the gluon distribution  given by the GRV98(LO)
parameterization \cite{grv98}. It should be noticed that different choices for the factorization scale and quark mass
produce distinct overall normalization to the total cross section
at photon-nucleon interactions and that NLO corrections can be absorbed in these  redefinitions of $\mu_F^2$ and $m_Q^2$. In the next section we discuss in more detail the dependence of our results in the parton distributions used as input in our calculations.

On the other hand, 
  in the large
energy (small-$x$) limit,  the effects of the finite transverse
momenta of the incoming partons become important, and the
factorization must be generalized, implying that the cross
sections are now $k_{\perp}$-factorized into an off-shell partonic
cross section and a  $k_{\perp}$-unintegrated parton density
function ${\cal{F}}(x,k_{\perp})$, characterizing the
$k_{\perp}$-factorization  approach \cite{CCH,CE,GLRSS}.  The
function $\cal{F}$ is obtained as a solution  of the   evolution
equations associated to the dynamics that governs the QCD at high
energies (see
\cite{smallx} for a review).  The latter can recover the usual
parton distributions in the double leading logarithmic limit (DLL) by its
integration over the transverse momentum of the $\rk_{\perp}$
exchanged gluon. The gluon longitudinal momentum fraction is
related to the c.m.s. energy, $W_{\gamma \,p}$,  in the heavy
quark photoproduction case as $x=4m_{Q}^2/W_{\gamma \,p}^2$. This assumption is a very good approximation, thought the scaling variable $x$ in fact depends on the kinematic variables  of the incoming particles, namely on $z$ and parton momenta (for details on these issues see \cite{Mariottomagno} and references therein). The cross section for the heavy-quark
photoproduction process is given by the convolution of the
unintegrated gluon function with the off-shell matrix elements
\cite{Mariottomagno,smallx,semih,Timneanu_Motyka}.  Considering
only the direct component of the photon we have that the total cross section  reads  as \cite{Mariottomagno},
\begin{eqnarray}
\sigma_{tot}\,(\gamma p\rightarrow Q\overline{Q}X) & = & 
  \frac{e_Q^2}{\pi}\, \int\, dz\,d^2 \rp_{1\perp} \frac{d^2\rk_{\perp}}{\rk_{\perp}^2}\, \alpha_{em}\,\alpha_s(\mu^2)\,{\cal F}(x,\rk_{\perp}^2;\,\mu^2)\,\nonumber \\
& \times &  \left\{[z^2+(1-z)^2]\,\left( \frac{\rp_{1\perp}}{D_1} + \frac{(\rk_{\perp}-\rp_{1\perp})}{D_2} \right)^2 +   m_Q^2 \,\left(\frac{1}{D_1} + \frac{1}{D_2}  \right)^2  \right\}\,, \label{sigmakt}
\end{eqnarray}
where $D_1 \equiv \rp_{1\perp}^2 + m_Q^2$ and $D_2 \equiv (\rk_{\perp}-\rp _{1\perp})^2 + m_Q^2$. The transverse momenta of the heavy quark (antiquark) are denoted by $\rp_{1\perp}$ and $\rp_{2\perp}= (\rk_{\perp}-\rp _{1\perp})$, respectively. The heavy quark longitudinal momentum fraction is labeled by $z$. The hard scale $\mu$ in general is taken to be equal to the gluon virtuality,  in close connection with the BLM scheme \cite{BLM}. Therefore, for our purpose we will use the prescription  $\mu^2=\rk_{\perp}^2 + m_{Q}^2$. As some evolution equations for the unintegrated gluon distribution, for instance the CCFM equation \cite{ccfm}, enable also its  dependence on the scale $\mu$, we have explicitly  written down it in Eq. (\ref{sigmakt}) for sake of generality.

In performing a numerical analysis within the
$k_{\perp}$-factorization approach (hereafter labeled SEMIHARD), we use two
distinct  parameterizations for the unintegrated gluon
distribution (For details see Ref. \cite{Goncalves:2003zy}).
First, one considers the derivative of the collinear gluon distribution, quite successful in the proton case \cite{Mariottomagno}
and investigated in the nuclear case in Ref. \cite{Goncalves:2003zy}. It
simply reads  as 
\begin{eqnarray}
{\cal F}_{\mathrm{DLL}}\,(x,\,\rk_{\perp}^2) =  \frac{\partial\, xg(x,\,\rk_{\perp}^2)}{\partial \ln \rk_{\perp}^2}\,,
\label{ugfdll}
\end{eqnarray}
where $xg(x,Q^2)$ is the gluon distribution, which was
taken from  the GRV98(LO)  parton distribution
\cite{grv98}. A  shortcoming of the function above is that it eventually produces negative values for the unintegrated gluon distribution at large $x$. This has consequences in the description of the behavior near threshold, being more important for bottom production than for charm. This is due to the scaling variable to be proportional to the quark mass, $x=4m_Q^2/W_{\gamma p}^2$, and in general the calculation slightly underestimates the cross section value at low energies. A more accurate calculation consists on  multiplying the collinear gluon function above by the Sudakov-like form factor $T_g(\rk,\,\mu)$ \cite{kimber}. 

As the heavy quark production is characterized by a scale of order of the quark mass, the charm production, in particular, is considered a primary candidate for investigating the kinematic region of QCD on the boundary between perturbative and high density QCD (For recent reviews see, e.g., Refs. \cite{raju_iancu,mpla}). Recently, we have demonstrated that the inclusive charm total cross section  exhibits the property of geometric scaling \cite{prl}, which is one of the main characteristics of the high density approaches. This fact  motivates to  estimate these features in the photoproduction case. In order to do so, we  consider the  Golec-Biernat-W\"{u}sthoff parameterization \cite{GBW} (hereafter SAT-MOD), which gives the following unintegrated gluon distribution,
\begin{eqnarray}
 {\cal F}_{\mathrm{sat}}\,(x,\,\rk_{\perp}^2)= \frac{3 \sigma_0 }{4\,\pi^2 \alpha_s\,}\,\left(\frac{\rk_{\perp}^2}{Q_{\mathrm{sat}}^2(x)} \right)\, \exp \left(-\frac{\rk_{\perp}^2}{Q_{\mathrm{sat}}^2(x)}  \right)\,(1-x)^7\,,\hspace{1cm} Q_{\mathrm{sat}}(x) =  \left( \frac{x_0}{x}
\right)^{\lambda} \,\,\mathrm{GeV}^2\,,
\label{gluonsat}
\end{eqnarray}
where $Q_{\mathrm{sat}}^2(x)$ is the saturation scale, which defines the onset of the nonlinear QCD (saturation) effects. The parameters were obtained from a fit of the inclusive structure function $F_2$ and total photoproduction cross section data at DESY-HERA data. The fit using the expression of the color dipole picture, as explained below, and the saturation ansatz for the dipole cross section produced  $\sigma_0=29.12$ mb, $\lambda= 0.277$ and $x_0=0.41 \cdot 10^{-4}$. This fit procedure  also includes  the charm contribution. In addition, we have included the large-$x$ threshold factor $(1-x)^7$ (for the heavy quark case) in Eq. (\ref{gluonsat}) to take into account the correct energy dependence at low energies near production threshold. 

The saturation model is based on the color dipole picture of the photon-proton interaction \cite{dipole}. In the proton rest frame, the DIS process can be seen as 
a succession in time of three factorisable subprocesses: i) 
the photon fluctuates in a quark-antiquark pair with transverse separation $\rr\sim 1/Q$ long after the interaction, ii) this 
color dipole interacts with the proton target, iii) the quark pair
annihilates in a virtual photon. The interaction $\gamma^*p$ is further
factorized in the simple formulation \cite{dipole},
\begin{eqnarray}
\sigma_{L,T}^{\gamma^*p}(x,Q^2)=\sum_f\,\int dz \,d^2\rr\,
|\Psi_{L,T}^{f}\,(z,\rr,Q^2)|^2
\,\sigma_{dip}(x,\rr)\,,
\end{eqnarray}
where $z$ is the longitudinal momentum fraction of the quark. The photon wavefunctions $\Psi_{L,T}^{f}$ are determined
from light cone perturbation theory and read as
\begin{eqnarray}
 |\Psi_{T}^f|^2 & = &\!  \frac{6\alpha_{\mathrm{em}}}{4\,\pi^2} \,
  e_f^2 \, \left\{[z^2 + (1-z)^2]\, \varepsilon^2 \,K_1^2(\varepsilon \,\rr)
 +\,  m_f^2 \, \,K_0^2(\varepsilon\,\rr)
 \right\}\label{wtrans} \nonumber \\
 |\Psi_{L}^f|^2 & = &\! \frac{6\alpha_{\mathrm{em}}}{\pi^2} \,
 e_f^2 \, \left\{Q^2 \,z^2 (1-z)^2
\,K_0^2(\varepsilon\,\rr) \right\}, \label{wlongs}
 \end{eqnarray}
where the auxiliary variable $\varepsilon^2=z(1-z)\,Q^2 + m^2_f$
depends on the quark mass, $m_f$. The $K_{0,1}$ are the McDonald
functions and the summation is performed over the quark flavors. The dipole-hadron cross section $\sigma_{dip}$  contains all
information about the target and the strong interaction physics. In general, the saturation models \cite{iancu_munier,GBW,bgbk,kowtea}  interpolate between the small and large dipole
configurations, providing color transparency behavior, $\sigma_{dip} \sim \rr^2$, as $\rr \gg Q_{\mathrm{sat}}$  and constant behavior at
large dipole separations $\rr < Q_{\mathrm{sat}}$.  Along these lines, the phenomenological saturation model resembles the mains features of the Glauber-Mueller resummation. Namely, the dipole cross section takes the eikonal-like form,
\begin{eqnarray}
\sigma_{dip} (x, \,\rr)  =  \sigma_0 \, \left[\, 1- \exp
\left(-\frac{\,Q_{\mathrm{sat}}^2(x)\,\rr^2}{4} \right) \, \right]\,. \label{gbwdip}
\end{eqnarray}
Its phenomenological application has been successful in a wide class of processes with a photon probe. An  important aspect of the saturation models is that they resume a class of higher twist contributions which should be non-negligible towards the low $Q^2$ region \cite{peters,flvicmag}. Moreover, it is important to emphasize that the dipole approach was extended for heavy quark production in proton-proton (nucleus) collisions in Refs. \cite{nikolaev,kopeliovich}, and  its equivalence with the collinear approach as well as the comparison with $pp$ data  was  presented in Ref. \cite{rauf}.

 Although the  saturation model describes reasonably well the  HERA data, its functional form is only an approximation of the theoretical non-linear QCD approaches. On the other hand, an analytical expression for the dipole cross section can be obtained within the BFKL formalism. Currently, intense theoretical studies has been performed towards an understanding of the BFKL approach in the border of the saturation region \cite{IANCUGEO,MUNIERWALLON}. In particular, the dipole cross section has been calculated in both LO  and NLO BFKL  approach in the geometric scaling region \cite{BFKLSCAL}. It reads as,
\begin{eqnarray}
\sigma_{dip}(x,\rr)=\sigma_0\,\left[\rr^2 Q_{\mathrm{sat}}^2(x)\right]^{\gamma_{\mathrm{sat}}}\,\exp\left[ -\frac{\ln^2\,\left(\rr^2 Q_{\mathrm{sat}}^2\right)}{2\,\beta \,\bar{\alpha}_sY}\right]\,,
\label{sigmabfkl}
\end{eqnarray}
where $\sigma_0  =2\pi R_p^2$ ($R_p$ is the proton radius) is the overall normalization and the power $\gamma_{\mathrm{sat}}$ is the (BFKL) saddle point in the vicinity of the saturation line $Q^2= Q_{\mathrm{sat}}^2(x)$ (the anomalous dimension is defined as $\gamma = 1- \gamma_{\mathrm{sat}}$). As usual in the BFKL formalism, $\bar{\alpha}_s=N_c\,\alpha_s/\pi$, $\beta \simeq 28\,\zeta (3)$ and  $Y=\ln (1/x)$. The quadratic diffusion factor in the exponential gives rise to the scaling violations.  

The dipole cross section  in Eq. (\ref{sigmabfkl}) does not include an extrapolation from the geometric scaling region to the saturation region. This  has been recently implemented in Ref. \cite{iancu_munier}, where the dipole amplitude  ${\mathcal N} (x,\rr)=\sigma_{dip}/2\pi R_p^2$ was constructed to smoothly interpol between the  limiting behaviors analytically under control: the solution of the BFKL equation
for small dipole sizes, $\rr\ll 1/Q_{\mathrm{sat}}(x)$, and the Levin-Tuchin law \cite{levin}
for larger ones, $\rr\gg 1/Q_{\mathrm{sat}}(x)$. A fit to the structure function $F_2(x,Q^2)$ was performed in the kinematical range of interest, showing that it is  not very sensitive to the details of the interpolation. The dipole cross section was parameterized as follows,
\begin{eqnarray}
\sigma_{dip}^{\mathrm{CGC}}\,(x,\rr) = \sigma_0\, \left\{ \begin{array}{ll} 
{\mathcal N}_0\, \left(\frac{\rr\, Q_{\mathrm{sat}}}{2}\right)^{2\left(\gamma_{\mathrm{sat}} + \frac{\ln (2/\rr Q_{\mathrm{sat}})}{\kappa \,\lambda \,Y}\right)}\,, & \mbox{for $\rr Q_{\mathrm{sat}}(x) \le 2$}\,,\\
 1 - \exp \left[ -a\,\ln^2\,(b\,\rr\, Q_{\mathrm{sat}}) \right]\,,  & \mbox{for $\rr Q_{\mathrm{sat}}(x)  > 2$}\,, 
\end{array} \right.
\label{CGCfit}
\end{eqnarray}
where the expression for $\rr Q_{\mathrm{sat}}(x)  > 2$  (saturation region)   has the correct functional
form, as obtained either by solving the Balitsky-Kovchegov (BK) equation \cite{BK}, 
or from the theory of the Color Glass Condensate (CGC) \cite{CGC}. Hereafter, we label the model above by CGC. The coefficients $a$ and $b$ are determined from the continuity conditions of the dipole cross section  at $\rr Q_{\mathrm{sat}}(x)=2$. The coefficients $\gamma_{\mathrm{sat}}= 0.63$ and $\kappa= 9.9$  are fixed from their LO BFKL values. In our further calculations it will be used the parameters $R_p=0.641$ fm, $\lambda=0.253$, $x_0=0.267\times 10^{-4}$ and ${\mathcal N}_0=0.7$, which give the best fit result. We have included also a large-$x$ factor as for the saturation model. In the color dipole picture, the heavy quark photoproduction cross section using the CGC model reads as,
\begin{eqnarray}
\sigma_{tot}\,(\gamma p \rightarrow Q\overline{Q}X)=\int dz \,d^2\rr \,
|\Psi_{T}^{Q}\,(z,\,\rr,Q^2=0)|^2
\,\sigma_{dip}^{\mathrm{CGC}}(x,\rr)\,,
\end{eqnarray}
where the longitudinal piece does not contribute as $|\Psi_{L}|^2\propto Q^2$. The transverse contribution is computed using Eq. (\ref{wtrans}) and introducing the appropriated mass and charge of the charm or  bottom quark.

The dipole approach is very useful in providing semi-analytic solutions for the cross section. In the heavy quark case, it can be shown that the scattering process is dominated by small size dipoles with mean value $\rr \sim 1/m_Q^2$. In this scenario, the photon wavefunction selects the color transparency region as the main contribution to the cross section. Concerning the longitudinal fraction $z$ of the dipoles,  at the limit  $Q^2\ll 4\,m_Q^2$ only the symmetric configurations $<\!z\!>\approx 1/2$ contribute. In this case, which it is obeyed in the photoproduction regime, we can write in a semi-quantitative way,
\begin{eqnarray}
\sigma_{tot}(\gamma p \rightarrow Q\overline{Q}X) \simeq \frac{\alpha_{\mathrm{em}}\,e_Q^2}{\pi}\,\int_0^{1/m_Q^2} \frac{d\rr^2}{\rr^2}\,\bar{\sigma}(x,\rr)\,,
\label{semianalytic}
\end{eqnarray} 
where $\bar{\sigma}$ is the small-$r$ limit of the dipole cross section. For heavy quark production we have that the saturation model predicts  $\bar{\sigma}=\sigma_0\,Q_{\mathrm{sat}}^2\,\rr^2/4$, whereas for the CGC model it reads as  $\bar{\sigma}=\sigma_0\,{\cal N}_0\,(Q_{\mathrm{sat}}^2\,\rr^2/4)^{\gamma_{\mathrm{sat}}}$ modulo the diffusion term in Eq. (\ref{CGCfit}). Therefore, by introducing these values  in Eq. (\ref{semianalytic}) one  produces the following  analytical results for both models,
\begin{eqnarray}
\sigma_{tot}^{\mathrm{SAT-MOD}}\simeq \frac{\alpha_{\mathrm{em}}\,e_Q^2}{\pi}\left(\frac{\sigma_0\,Q_{\mathrm{sat}}^2}{4\,m_Q^2} \right) \propto W_{\gamma p}^{2\lambda}\,,
\hspace{0.5cm} \sigma_{tot}^{\mathrm{CGC}}\simeq \frac{\alpha_{\mathrm{em}}\,e_Q^2\,\sigma_0 \,{\cal N}_0}{\pi\,m_Q^2\,(1+\gamma_{\mathrm{sat}})}
\left(\frac{Q_{\mathrm{sat}}^2}{4\,m_Q^2}\right)^{\gamma_{\mathrm{sat}}}\propto W_{\gamma p}^{2\gamma_{\mathrm{sat}}\lambda}\,,
\label{sigtot_analytic}
\end{eqnarray}
which implies in general a slightly smoother energy growth for CGC than for SAT-MOD since $\gamma_{\mathrm{sat}}<1$. It should be noticed that this behavior is  obtained only in the case of sufficiently small $x$ (large rapidity Y), where we could neglect the diffusion term for CGC and take only the anomalous dimension at the saturation vicinity. On the other hand, for not so small  $x$ the diffusion factor can not be disregarded and the effective anomalous dimension $\gamma_{\mathrm{eff}}\,(x,\rr)=  \gamma_{\mathrm{sat}} + [\,\ln \,(2/\rr Q_{\mathrm{sat}})/\kappa \,\lambda \,Y \,]$ in Eq. (\ref{CGCfit}) should be taken. This situation can occur in the bottom production at intermediate energies due to the large bottom mass, which implies the CGC having practically the same slope on energy as the saturation model. This can be understood in realizing that  the effective anomalous dimension for SAT-MOD is $\gamma_{\mathrm{eff}}^{\mathrm{SAT-MOD}}=1$ and at sufficiently large $x$ they are very close of each other.

 Having presented the main sound approaches for heavy quark photoproduction at the photon level, let us compare their numerical results with the experimental DESY-HERA data \cite{h1data}. They are shown in Fig. \ref{fig1}, where the following notation is considered: the results from the saturation model (SAT-MOD) are denoted by the solid lines and the usual collinear factorization calculation is labeled by the dot-dashed curves. The $k_{\perp}$-factorization formalism (SEMIHARD), using the unintegrated gluon function in Eq. (\ref{ugfdll}) produces the long-dashed curves, where the Iancu-Itakura-Munier parameterization for the dipole cross section (CGC) gives the dashed curves. In all calculations we have use the same quark masses $m_c = 1.5$ GeV and $m_b = 4.5$ GeV. We quote Refs. \cite{Mariottomagno,Goncalves:2003zy} for detailed investigations on the dependence of the cross section in choosing different collinear gluon parameterizations and different factorization scale for both the collinear and $k_{\perp}$-factorization approaches.

Lets compare the distinct behaviors on energy and overall normalizations. Both the collinear approach and the semihard formalism give a consistent data description at high energies. At the region near the threshold, the semihard approach slightly underestimate the cross section, being more pronounced in the bottom case. This is understood in terms of the considered unintegrated gluon distribution ${\cal F}_{\mathrm{DLL}}$, as discussed before. The saturation model underestimates the cross section at high energies by a factor about 2, producing a reasonable description of the region near threshold. The CGC model follows similar trends, but the high energy values are closer to the experimental measurements than the saturation model. The slope on energy for SAT-MOD and CGC can be easily understood from the semi-quantitative results in Eq. (\ref{sigtot_analytic}). For the bottom case, we are not in sufficiently small $x$ and SAT-MOD and CGC give similar energy slope in agreement with our discussion above. Unfortunately, the current precision and statistics of the experimental measurements of the photoproduction cross section are either low to formulate  definitive conclusions about the robustness of the different approaches presented here. More precise measurements could be pose stringent constraints on the energy dependence and overall normalization. Finally, it should be noticed that the present calculations concern only the direct photon contribution to the cross section, whereas the resolved component has been neglected. In some extent the results from the saturation models presented here let some room for this contribution. Details on its calculation and size of its contribution can be found, for instance,  in Ref. \cite{Timneanu_Motyka}.

\section{Results and conclusions}

In what follows, we will compute the rapidity distribution and total cross sections for the photoproduction of open charm and bottom from proton-proton collisions at high energies. The approaches shortly reviewed in the previous section serve as input for the numerical calculations using Eq. (\ref{eq:sigma_pp}) for  the energies of the  current and future $pp$ and $p\bar{p}$ accelerators.  Namely, one considers the shorter $pp$ running at RHIC upon energy of $\sqrt{S_{NN}}=500$ GeV  and  the Tevatron value $\sqrt{S_{NN}}=1.96$ TeV for its $p\bar{p}$ running.  For the planned LHC $pp$ running (ATLAS and/or CMS) one takes the design energy  $\sqrt{S_{NN}}=14$ TeV.

The distribution on rapidity $y$ of the produced open heavy quark state can be directly computed from Eq. (\ref{eq:sigma_pp}), by using its  relation with the photon energy $\omega$, i.e. $y\propto \ln \, (\omega/m_Q)$.  A reflection around  $y=0$ takes into account the interchanging between the proton's photon emitter and the proton target. Explicitly, the rapidity distribution is written down as,
\begin{eqnarray}
\frac{d\sigma \,\left[p+p(\bar{p}) \rightarrow Q\overline{Q} + X + p(\bar{p} ) \right]}{dy} = \omega \, \frac{N_{\gamma} (\omega )}{d\omega }\,\sigma_{\gamma p \rightarrow Q\overline{Q}X}\,\left(\omega \right)\,.
\end{eqnarray}

Lets start by  analyzing the predictions from the collinear approach, obtained using distinct gluon pdf's. In Fig. \ref{figcomparacao}, the comparison among  the resulting rapidity distributions for open charm photoproduction at LHC  energy is presented. It has been considered the  GRV (94 and 98) and MRST (2001) parton parameterizations. The results for bottom and/or charm  at RHIC/Tevatron energies give somewhat very close curves and they will be not presented here. The results presented in  Fig. \ref{figcomparacao} are strongly dependent on the gluon distribution, which motivates the study of this process in order to constrain it. Moreover, they can useful in understanding what is the underlying QCD dynamics at high energies. We have that the deviations between the predictions from the GRV98(LO) and MRST2001(LO) pdf's are  approximately of 30 \%.  In order to demonstrate the strong dependence of the rapidity distribution on the gluon density, we also present the result obtained from the obsolete GRV94(LO) pdf, which it has a steeper gluon distribution at small $x$. In our further comparisons with other approaches, we will use the GRV98 pdf. 

\begin{figure}[t]
\begin{tabular}{cc}
\includegraphics[scale=0.5] {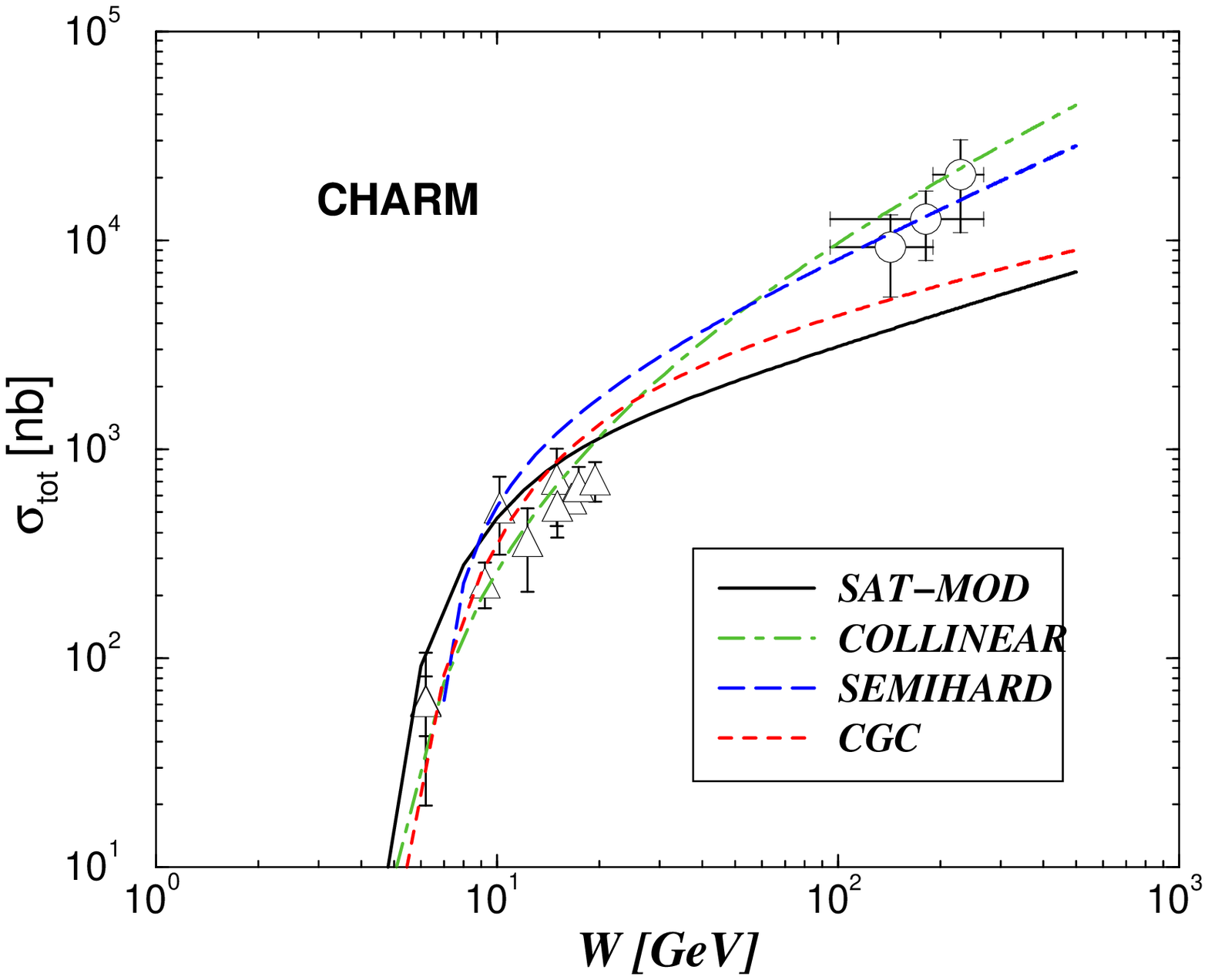} & \includegraphics[scale=0.5]{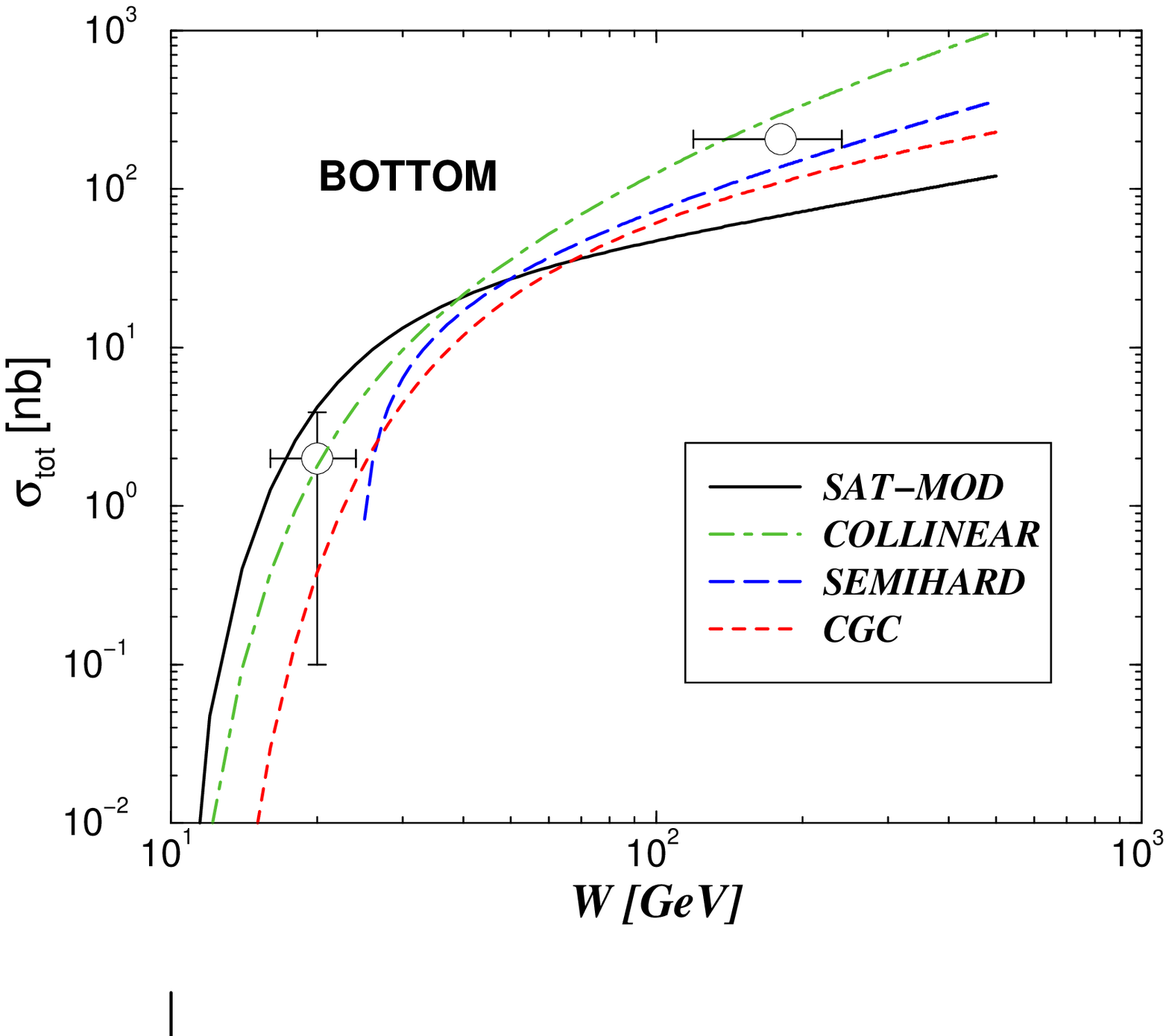}
\end{tabular}
\caption{\it The total photoproduction cross section for charm (left panel) and bottom (right panel). The experimental measurements are from DESY-HERA. }
\label{fig1}
\end{figure}

\begin{figure}[t]
\includegraphics[scale=0.5]{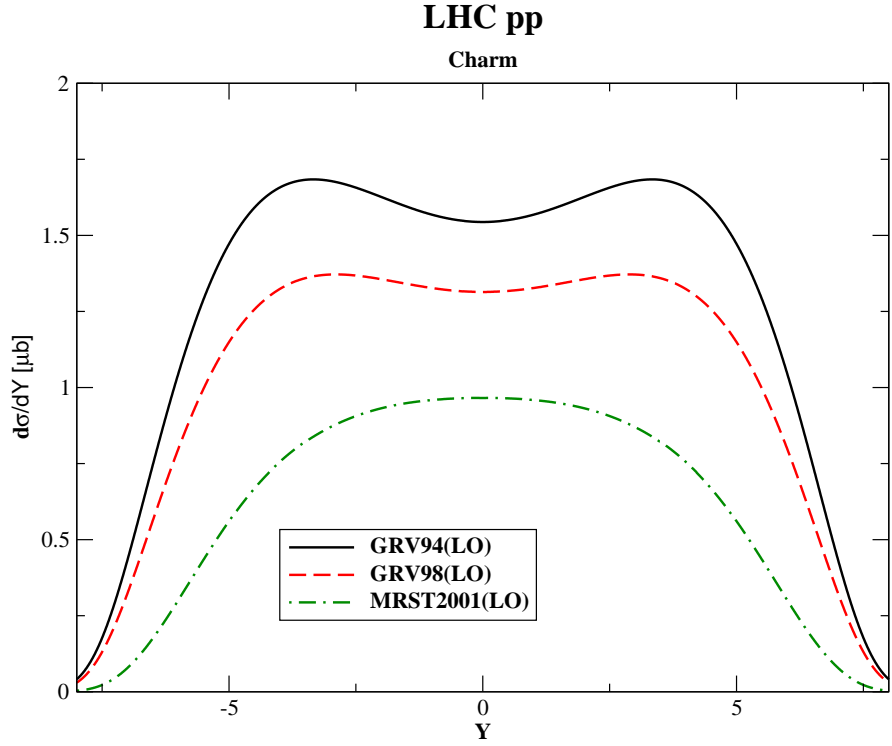}
\caption{\it Comparison among rapidity distributions for charm photoproduction at LHC,  considering distinct parameterizations for the collinear gluon distribution.}
\label{figcomparacao}
\end{figure}

\begin{figure}[t]
\begin{tabular}{cc}
\includegraphics[scale=0.5] {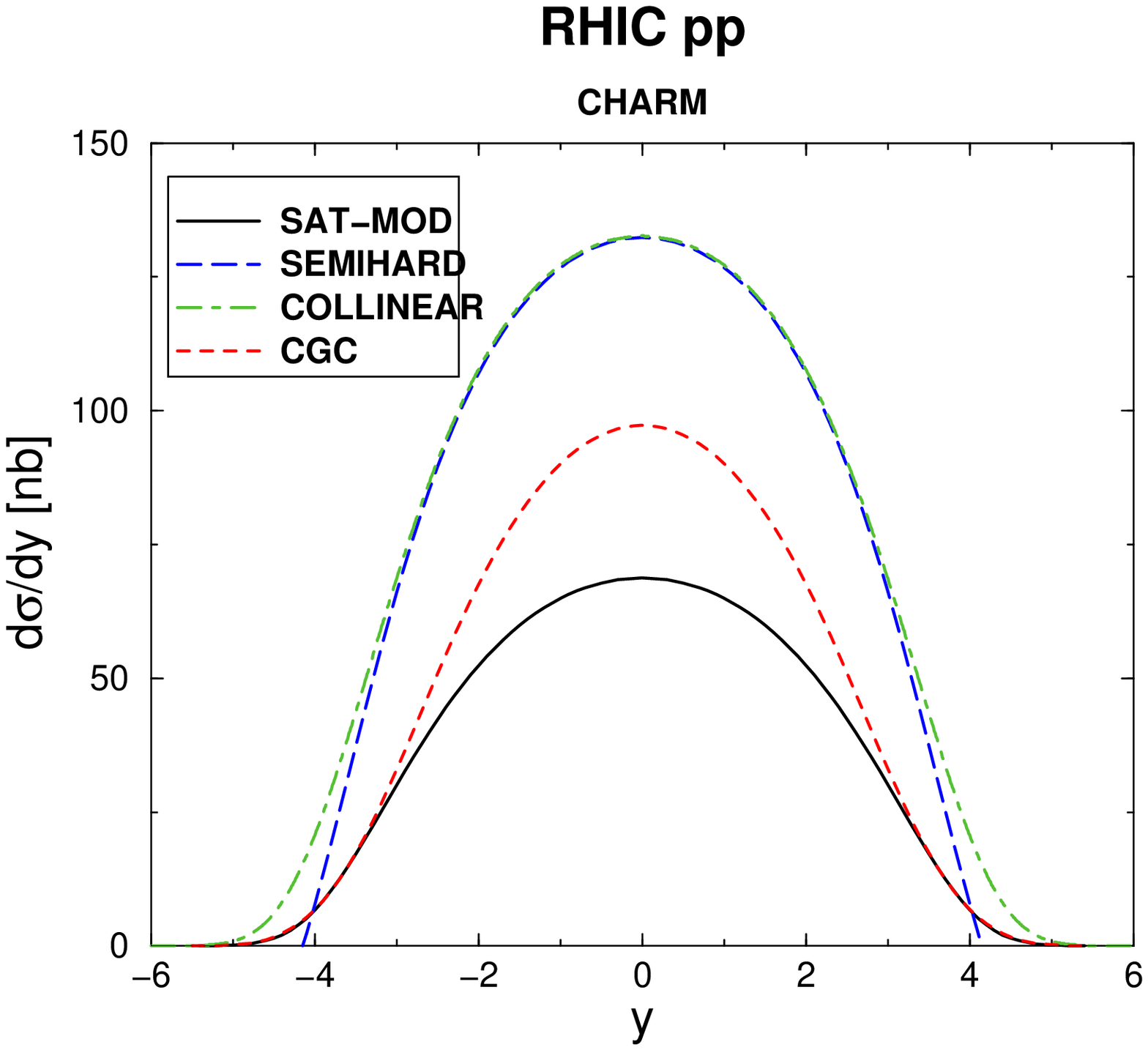} & \includegraphics[scale=0.5]{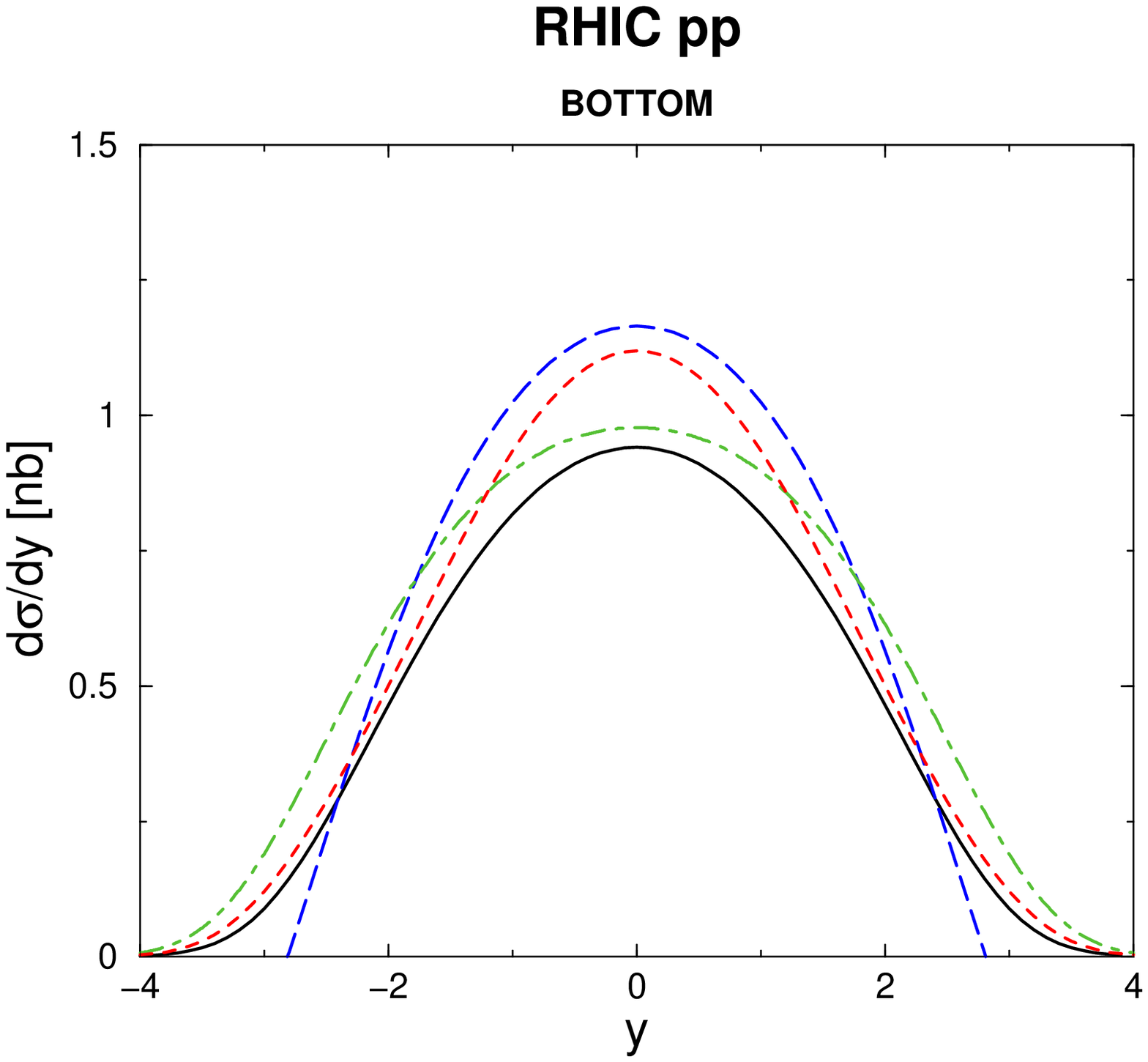}
\end{tabular}
\caption{\it The rapidity distribution for open charm and bottom photoproduction on $pp$ reactions at RHIC energy $\sqrt{S_{NN}}=500\,\,\mathrm{GeV}$. Different curves correspond to distinct high energy QCD approaches (see text).}
\label{fig2}
\end{figure}

\begin{figure}[t]
\begin{tabular}{cc}
\includegraphics[scale=0.5] {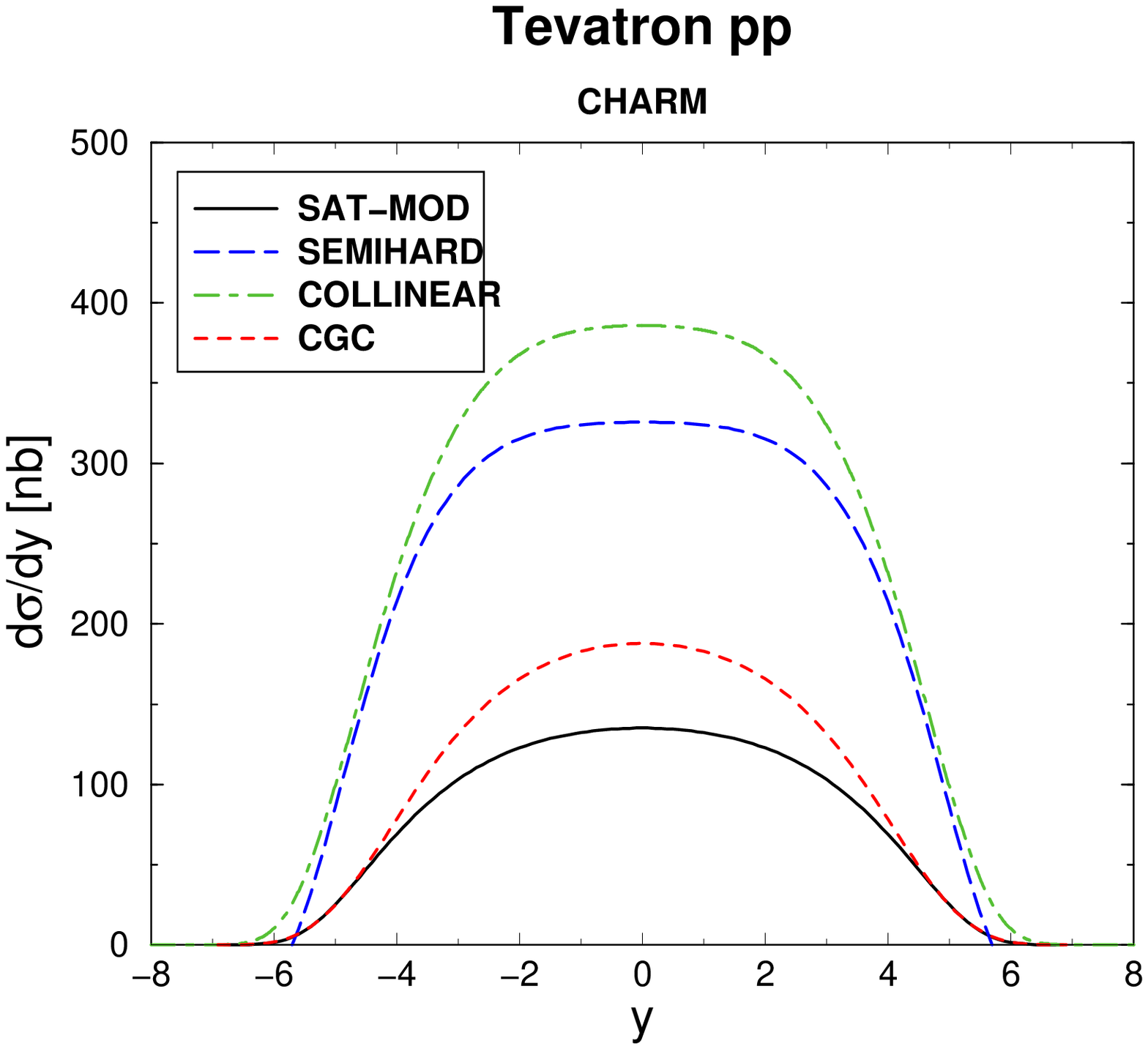} & \includegraphics[scale=0.5]{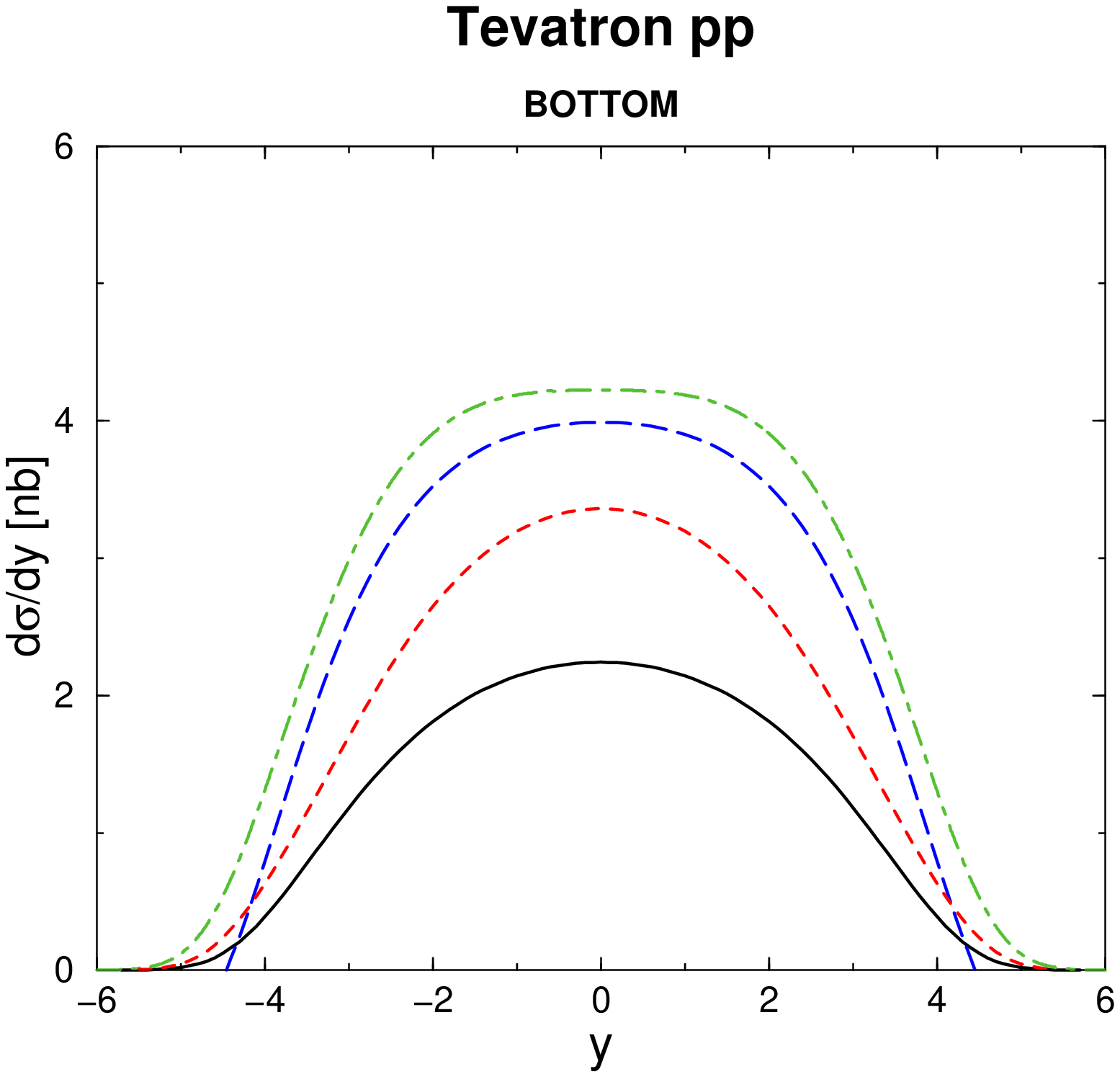}
\end{tabular}
\caption{\it The rapidity distribution for open charm and bottom photoproduction on $p\bar{p}$  reactions at Tevatron  energy $\sqrt{S_{NN}}=1.96\,\,\mathrm{TeV}$. Different curves correspond to distinct high energy QCD approaches (see text).}
\label{fig3}
\end{figure}

\begin{figure}[t]
\begin{tabular}{cc}
\includegraphics[scale=0.5] {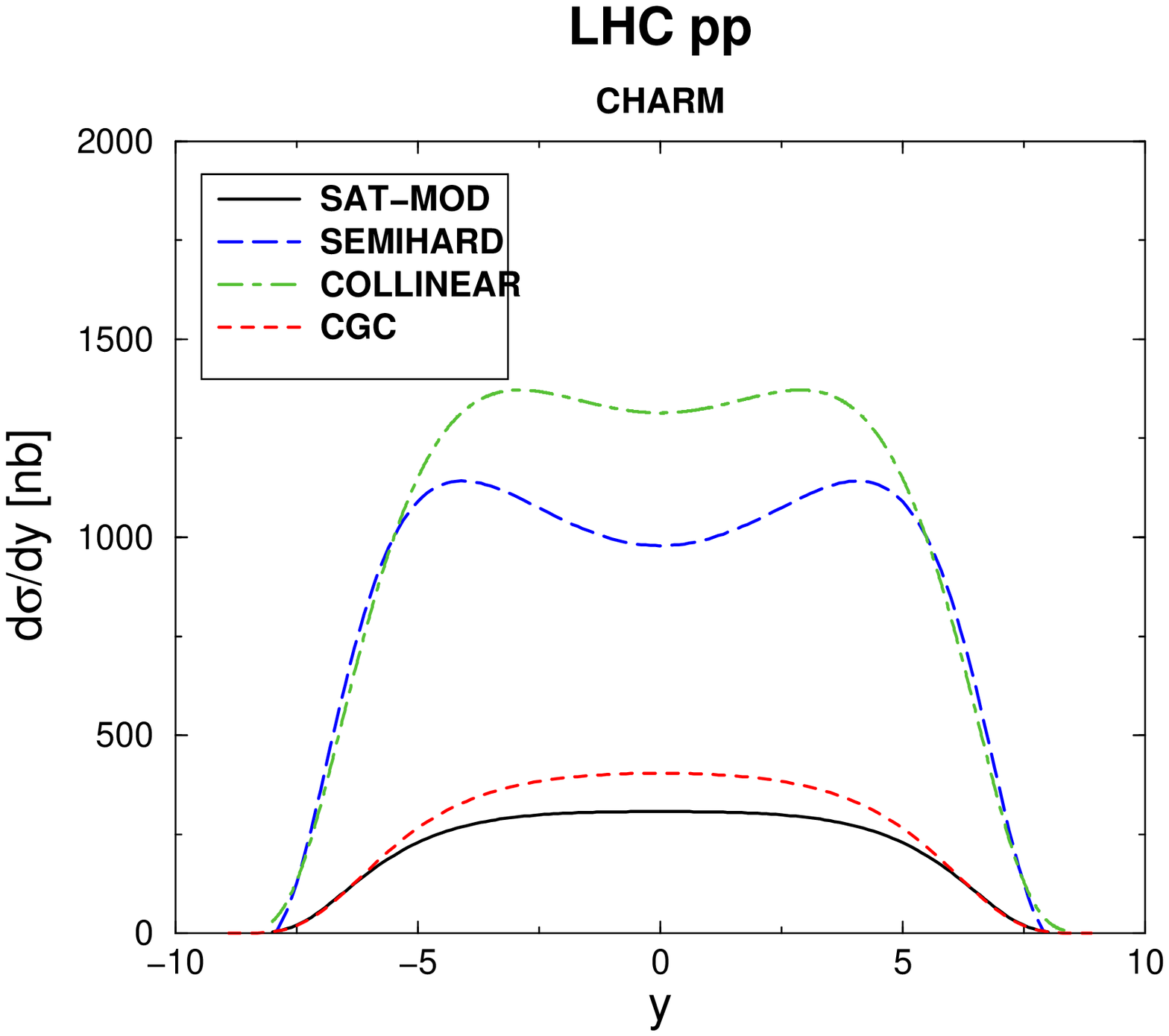} & \includegraphics[scale=0.5]{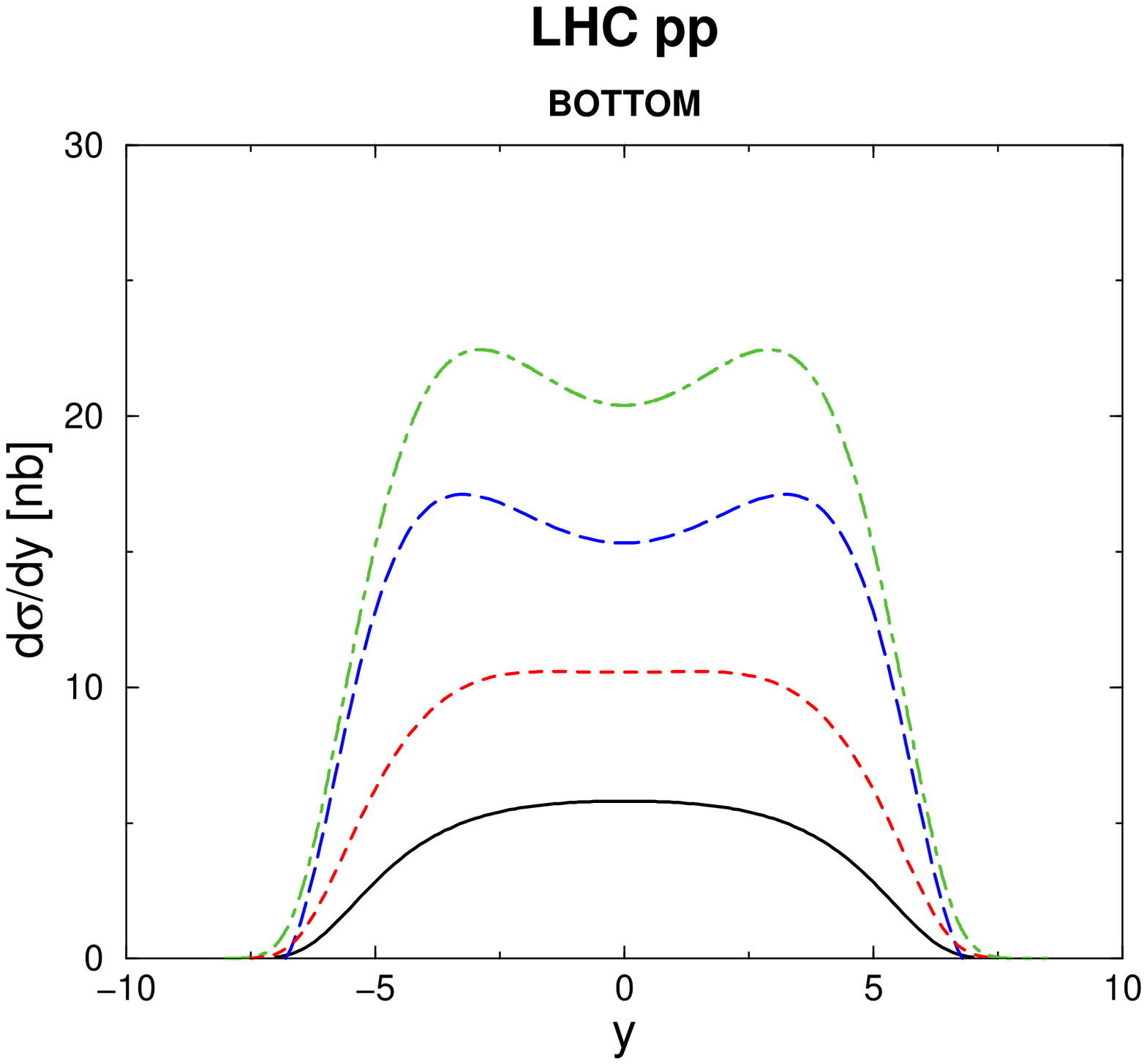}
\end{tabular}
\caption{\it The rapidity distribution for open charm and bottom photoproduction on $p\bar{p}$  reactions at LHC (CMS and/or ATLAS)  energy $\sqrt{S_{NN}}=14\,\mathrm{TeV}$. Different curves correspond to distinct high energy approaches (see text).}
\label{fig4}
\end{figure}

The resulting cross sections  coming out of the distinct theoretical inputs considered in previous section are depicted in Figs.  (\ref{fig2}-\ref{fig4}) at  RHIC, Tevatron and LHC energies, respectively. Lets discuss in general lines the present results. At RHIC energies, larger values of $x\propto m_Q/W_{\gamma p}$ are probed in the process than at Tevatron and LHC, mostly for bottom. This fact explains the very similar result for all approaches in the bottom case in Fig. \ref{fig2} (right plot). For the charm case, the situation starts to be different. The SEMIHARD result presents a sharper tail at large rapidities as a consequence of its behavior on energy near threshold, as discussed before. The overall normalization and its behavior at central rapidities follow the same trend coming from the cross section at photon level (see Fig. \ref{fig1}). Collinear and $k_{\perp}$-factorization formalisms often give closer results, whereas the saturation models (SAT-MOD and CGC) produce smaller cross sections.

At Tevatron energies (Fig. \ref{fig3}), smaller $x$ are being probed in the reaction, and the cross section becomes more dependent on the high energy behavior of the cross section at photon level. The overall normalization becomes increasingly sizeable for different approaches. For the charm case, the result could already distinguish between the saturation approaches and the usual pQCD calculations. At LHC energies (Fig. \ref{fig4}), the separation is more clear once smaller $x$ are probed. High values are obtained from the collinear approach  due to its steeper energy growth, followed by the semihard approach. The latter still considers an unintegrated gluon function dependent on the collinear one and therefore has closer behavior on energy.  Concerning the saturation approaches, SAT-MOD provides the lower limit, whereas CGC gives somewhat larger values. This is once again a consequence of their high energy behavior, as verified in Fig. \ref{fig1}.

Let us now  compute the integrated  cross section considering the
distinct QCD approaches. The results are presented in Table \ref{tabhq}, 
for the open charm and bottom pair production at RHIC, Tevatron and LHC, respectively. The collinear factorization approach gives
the  largest rates among the models studied, followed by the semihard formalism, as a clear trend
from the distribution on rapidity. Concerning the saturation models, the CGC results in a cross section of order 20 \% larger than SAT-MOD. The values are either large at Tevatron and LHC, going from some units of $\mu$bs at RHIC  to hundreds of $\mu$bs at LHC. Therefore, these reactions can have high rates at the  LHC kinematical regime. As stated before, we don't consider the resolved photon contribution in our calculations. However, it is possible to present an estimate of this process considering the results from Refs. \cite{klein_vogt,vic_bert} for ultraperipheral heavy ion collisions, where this contribution was studied using collinear factorization. One of the  main results is that these contributions are $\approx 15 \%$ and $20 \%$ of the total charm and bottom photoproduction cross sections at LHC energy, respectively. On the other  hand, in Ref. \cite{Timneanu_Motyka} this contribution was estimated to be of order of 20-30 $\%$  of the direct photon cross section using $k_{\perp}$-factorization. It is important to emphasize that the inclusion of resolved photon contribution brings the predictions for bottom production closer to the $\gamma p$ data, but a major inconsistency cannot be claimed due to the large experimental errors and theoretical uncertainties.  We postpone a  detailed analysis on these issues for a future publication. 

The cross section computed here can be contrasted with the open charm hadroproduction in central $pp(\bar{p})$ collisions. Recently, Raufeisen and Peng \cite{rauf} have computed it considering the NLO parton model and the color dipole formulation. They found that the results are subject to uncertainties coming from different choices for quark mass and parameters of the models, but they are able to describe all data (except recent STAR measurement) inside the uncertainty band. The photoproduction cross section is of order 0.1 \% from the corresponding hadroproduction. However, as we will see below, the production rates are either high and the experimental signal is significantly clear. 

At RHIC, where the luminosity is assumed to be  ${\cal L}_{\mathrm{RHIC}} = 10^{31}$ cm$^{-2}$s$^{-1}$, the open charm rate ranges on $3.7-7.8\times 10^6$ events by year. It should be noticed that RHIC uses most part of the running in the heavy-ion mode, so we used $10^6$ s in the last estimation. For the bottom case, the rates are $3.7-4.3\times 10^{4}$ events by year. At Tevatron, assuming the the running time $10^7$ s and design luminosity ${\cal L}_{\mathrm{Tevatron}} = 2\times \,10^{32}$ cm$^{-2}$s$^{-1}$, we have for charm $2-6 \times 10^{9}$ and for bottom $2-6\times 10^{7}$ events/year. The LHC produces the greatest rates (${\cal L}_{\mathrm{LHC}} = 10^{34}$ cm$^{-2}$s$^{-1}$), giving for charm $3.5-17 \times 10^{10}$ and for bottom $5.5-24 \times 10^{8}$ events/year. Notice the large rate for bottom at LHC.

 Finally, lets discuss the experimental separation of this reaction channel. As emphasized in Ref. \cite{Kleinpp}, although the photoproduction cross section to be  a small fraction of the hadronic cross section, the separation of this channel is feasible if we impose the presence of a rapidity gap in the final state. It occurs due to the proton which is the photon emitter remains intact in the process. Similarly to the $J/\Psi$ case, we expect that a cut in the transverse momentum of the pair could eliminate most part of the  contribution associated to the hadroproduction of heavy quarks. 
 Moreover, in comparison with the hadroproduction of heavy quarks, the event multiplicity for photoproduction interactions is lower, which implies that it may be used as a separation factor between these processes. As stated in Ref. \cite{klein_vogt}, one way to select photoproduction events is to eliminate events where both protons breakup. This should eliminate almost all of the hadroproduction events while retaining most of the photoproduction interactions. In Ref. \cite{klein_vogt} the rejection factor $R$, which is the probability of finding a rapidity gap with width $y$ in a $pp$ collision has been calculated. For photoproduction, the authors have obtained $R = 0.04 \, (0.005)$ at RHIC (LHC), requiring a single rapidity gap and $y = 2$ (For details see Section VI from Ref. \cite{klein_vogt}). These  estimates can be directly applied  in our analysis. As photoproduction always leads to a rapidity gap, this requiriment should reject relatively few signal events, leading to a good signal to noise ratio for selecting these events.
An important background which it should be analyzed is the diffractive heavy quark production in the single diffraction process \cite{heyssler}. We postpone this study for a future publication.

In summary, we have computed the cross sections for photoproduction of open heavy quarks in $pp$ and $p\bar{p}$ collisions. This has been performed using  well
established QCD approaches, namely  the collinear and semihard
factorization formalisms  as well as  saturation models within the color dipole approach. For the
first time, quantitative predictions for these approaches
are presented.  The obtained values are shown to be sizeable  at the current accelerators energies (RHIC and Tevatron) and are  increasingly larger at LHC. The feasibility of detection of these reactions is encouraging, since their experimental signature should be suitably clear.  Furthermore, they enable to constraint already in the current colliders the
QCD dynamics since the main features from photon-proton collisions
hold in proton-proton collisions.

\begin{acknowledgments}
The authors are grateful to Professor Bernd Kniehl ({\it II. Institut f\"ur Theoretische Physics, Universit\"at Hamburg}) for calling our attention to the Ref. \cite{kniehl} and for his comments related to the equivalent - photon distribution.
One of us (M.V.T.M.) thanks the support of the High Energy  Physics
Phenomenology Group at the Institute of Physics, GFPAE IF-UFRGS,
Porto Alegre. This work was partially financed by the Brazilian
funding agencies CNPq and FAPERGS.
\end{acknowledgments}

\newpage

\begin{table}[t]
\begin{center}
\begin{tabular} {||c|c|c|c|c|c||}
\hline
\hline
& $Q\overline{Q}$   & {\bf SAT-MOD} & {\bf SEMIHARD} & {\bf COLLINEAR } & {\bf CGC}  \\
\hline
\hline
 {\bf RHIC} & $c\bar{c}$ &  377 nb & 687 nb & 782 nb &  492 nb \\
\hline
 & $b\bar{b}$ &  3.7 nb &  3.6 nb & 4.3 nb & 4.2 nb \\
\hline
\hline
{\bf Tevatron} & $c\bar{c}$ &  1.04 $\mu$b & 2.77 $\mu$b & 3.21 $\mu$b &  1.36 $\mu$b \\
\hline
  & $b\bar{b}$ &  13.3 nb &  24.0  nb & 29.2 nb & 19.8 nb \\
\hline
\hline
 {\bf LHC} & $c\bar{c}$ &  3.54 $\mu$b & 14.2 $\mu$b & 16.7 $\mu$b &  4.37 $\mu$b \\
\hline
&  $b\bar{b}$ &  55.0 nb &  182  nb & 236 nb & 108 nb \\
\hline
\hline
\end{tabular}
\end{center}
\caption{\it The integrated cross section for the photoproduction of heavy quarks in $pp(\bar{p})$  collisions at RHIC, Tevatron and LHC.}
\label{tabhq}
\end{table}

\newpage

\end{document}